\documentclass[journal, 10pt]{IEEEtran}

\usepackage{multirow}
\usepackage[font=scriptsize,caption=false,labelsep=space]{subfig}
\usepackage{mathrsfs}
\usepackage{graphicx,float,wrapfig,epstopdf,amsmath}
\usepackage[]{algorithmicx}
\usepackage{algpseudocode,algorithm}
\usepackage{balance}
\usepackage{mathtools,lipsum}
\usepackage{amsmath}
\usepackage{amssymb}
\usepackage{amsthm}
\usepackage{graphicx}
\usepackage{epstopdf}
\usepackage{times}
\usepackage{textcomp,cite}

\usepackage{url}
\usepackage{hyperref}

\usepackage{multirow}
\usepackage{threeparttable}
\graphicspath{{./figures/}}

\DeclareMathOperator*{\argmax}{argmax} 

\usepackage[table]{xcolor}
\definecolor{intnull}{RGB}{213,229,255}
\definecolor{inteins}{RGB}{128,179,255}
\definecolor{color1}{RGB}{199,209,232}
\definecolor{color2}{RGB}{230,231,233}

\newtheorem{theorem}{Theorem}

\newtheorem{proposition}[theorem]{Proposition}
\newtheorem{corollary}{Corollary}

\begin{document}

	\title{ 
		BSA-OMP: Beam-Split-Aware Orthogonal Matching Pursuit for THz Channel Estimation
	}

	\author{\IEEEauthorblockN{Ahmet M. Elbir, \textit{Senior Member, IEEE}, 
			and  Symeon Chatzinotas, \textit{Fellow, IEEE}
		}
		\thanks{This work was supported in part by the HORIZON Project TERRAMETA and the ERC Project AGNOSTIC.}
				\thanks{A. M. Elbir is with the
					Interdisciplinary Centre for Security, Reliability and Trust (SnT) at the University of Luxembourg, Luxembourg; and Duzce University, Duzce, Turkey, (e-mail: ahmetmelbir@gmail.com).} 
		%
		\thanks{S. Chatzinotas is with the SnT at the University of Luxembourg, Luxembourg (e-mail: symeon.chatzinotas@uni.lu). }
	}
	\maketitle
	
	\begin{abstract}
		Terahertz (THz)-band has been envisioned for the sixth generation wireless networks thanks to its  ultra-wide bandwidth and very narrow beamwidth. Nevertheless, THz-band transmission faces several unique challenges, one of which is beam-split which occurs due to the usage of subcarrier-independent analog beamformers and causes the generated beams at different subcarriers split, and point to different directions. Unlike the prior works dealing with beam-split by employing additional complex hardware components, e.g., time-delayer networks, a beam-split-aware orthogonal matching pursuit (BSA-OMP) approach is introduced to efficiently estimate the THz channel and  beamformer design without any additional hardware. Specifically, we design a BSA dictionary comprised of beam-split-corrected steering vectors which inherently include the effect of beam-split so that the proposed BSA-OMP solution automatically yields the beam-split-corrected physical channel directions. Numerical results demonstrate the superior performance of BSA-OMP approach against the existing state-of-the-art techniques.
	\end{abstract}
	\begin{IEEEkeywords}
		Terahertz, beam split, channel estimation, orthogonal matching pursuit, support recovery.
	\end{IEEEkeywords}
	%
	

	\section{Introduction}
	\label{sec:Introduciton}
	Terahertz (THz) band is expected to be a key component of the sixth generation  (6G) of wireless cellular networks because of its abundant available bandwidth. In particular, THz-empowered systems are envisioned to demonstrate revolutionary enhancement in the data rate ($>100\text{Gb/s}$), extremely low latency ($<1\text{ms}$) and ultra reliability ($99.999\%$)~\cite{thz_Rappaport2019Jun}.
	
	Although demonstrating the aforementioned advantages, signal processing  in THz band faces several THz-specific challenges that should be taken into account accordingly. These challenges include, among others, severe path loss due to spreading loss and molecular absorption, extremely-sparse path model, very short transmission distance and beam-split (see the full list in~\cite{elbir2022Aug_THz_ISAC,ummimoTareqOverview}). In order to combat some of these challenges, e.g., path loss, analogues to massive multiple-input multiple-output (MIMO) arrays in millimeter-wave (mm-Wave) systems~\cite{heath2016overview}, ultra-massive MIMO  architectures are envisioned, wherein subcarrier-independent analog beamformers are employed. Therefore, the directions of the generated beams at different subcarriers differentiate and point to different directions causing \textit{beam-split} phenomenon since the analog beamformers can only be designed with respect to a single subcarrier~\cite{thz_beamSplit}.  In  mm-Wave band ($0.03-0.3$ THz), however, the subcarrier frequencies are relatively closer, and \textit{beam-squint} is broadly used to described the same effect~\cite{beamSquint_FeiFei_Wang2019Oct,elbir2021JointRadarComm}. In comparison, the main lobes of the array gain corresponding to the lowest and highest subcarriers totally split and do not overlap  while the squinted beam can still cover the entire bandwidth~\cite{spatialWidebandWang2018May}.  For instance, the amount of beam-split is approximately $6^\circ$ ($0.4^\circ$) for $0.3$ THz with $30$ GHz ($60$ GHz with $1$ GHz) bandwidth, respectively for a broadside target (see, e.g., Fig.~\ref{fig_ArrayGain})~\cite{elbir2022Aug_THz_ISAC}.
	
	Existing solutions to mitigate beam-split can be categorized into two classes: Hardware-based~\cite{dovelos_THz_CE_channelEstThz2,thz_beamSplit} and algorithmic methods~\cite{thz_channelEst_beamsplitPatternDetection_L_Dai,elbir2022Jul_THz_CE_FL,elbir_THZ_CE_ArrayPerturbation_Elbir2022Aug,spatialWidebandWang2018May,widebandCE2_GonzalezComa2020Feb}. The first category of solutions mainly consider employing additional hardware components such as time-delayer networks (TDNs) similar to the works studied to mitigate beam-squint~\cite{beamSquint_FeiFei_Wang2019Oct}. The use of TDN allows the generation of a virtual beampattern to realize  a subcarrier-dependent analog beamformer to eliminate the impact of beam-split. Prior works on TDNs include the application of  channel estimation~\cite{dovelos_THz_CE_channelEstThz2} and hybrid precoder/combiner~\cite{dovelos_THz_CE_channelEstThz2,thz_beamSplit} design problems. The second category, on the other hand, involves advanced signal processing techniques such as beam-split pattern detection (BSPD)~\cite{thz_channelEst_beamsplitPatternDetection_L_Dai}, machine learning (ML)~\cite{elbir2022Jul_THz_CE_FL} and sparse Bayesian learning (SBL)~\cite{elbir_THZ_CE_ArrayPerturbation_Elbir2022Aug}. In comparison, TDN-based approaches are cost-demanding, i.e., a single TD consumes approximately $100$ mW, which is more than that of a phase shifter ($40$ mW) in THz~\cite{elbir2022Aug_THz_ISAC}. \textcolor{black}{Furthermore,  algorithmic techniques suffer from inaccurate support estimation~\cite{thz_channelEst_beamsplitPatternDetection_L_Dai,elbir2022Jul_THz_CE_FL}, low angular precision~\cite{thz_channelEst_beamsplitPatternDetection_L_Dai,spatialWidebandWang2018May} and high computational complexity~\cite{elbir_THZ_CE_ArrayPerturbation_Elbir2022Aug,spatialWidebandWang2018May,widebandCE2_GonzalezComa2020Feb}.}
	
	\textcolor{black}{In this work, we propose an orthogonal matching pursuit (OMP)-based approach for joint THz channel and beam-split estimation. The proposed approach is advantageous since it does not require an additional hardware architecture, e.g., TDNs, and computationally-efficient because it automatically mitigates the impact of beam-split, hence called beam-split-aware OMP (BSA-OMP).} In the proposed approach,  beam-split is treated as an array imperfection as inspired from~\cite{elbir_THZ_CE_ArrayPerturbation_Elbir2022Aug}, and the beam-split-corrupted array data is transformed to beam-split-free steering vectors. Using  array imperfection model, a novel BSA dictionary which inherently includes the effect of beam-split is constructed. The key idea of the BSA-OMP technique is that the amount of beam-split is known proportionally prior to THz channel estimation task, while beam-split depends on the unknown user locations. Using this observation, we construct a BSA dictionary as the set of beam-split-corrected steering vectors to generate an accurate beamspace spectra. Thus, the proposed BSA approach automatically mitigates the effect of beam-split for both channel estimation and beamforming. \textcolor{black}{Similar array transformation approaches have been previously investigated for mm-Wave in~\cite{spatialWidebandWang2018May,widebandCE2_GonzalezComa2020Feb}. \cite{spatialWidebandWang2018May} proposed an angular-delay rotation (ADR) method, which suffers from coarse beam-split estimation and high training overhead due to the use of complete discrete Fourier transform (DFT) matrix. Also, a subspace-based approach is considered in~\cite{widebandCE2_GonzalezComa2020Feb}, which requires high training data snapshots and it may not always be applicable for THz channels in case of the highly correlated received paths. Compared to the previous works~\cite{thz_channelEst_beamsplitPatternDetection_L_Dai,spatialWidebandWang2018May,widebandCE2_GonzalezComa2020Feb}, the BSA-OMP exhibits lower computational complexity and channel overhead while accurately mitigating the beam-split.} 
	
	

	%


	\section{Wideband THz MIMO Transceiver Model}
	\label{sec:probForm}
	Consider a wideband THz MIMO architecture with hybrid analog/digital beamforming over $M$ subcarriers. {\color{black}We assume that the base station (BS) has $N$ antennas and $N_\mathrm{RF}$ radio-frequency (RF) chains to serve $K$ $\bar{N}$-antenna users. Let  $\mathbf{s}[m] = [s_1[m],\cdots,s_K[m]]^\textsf{T}$ be the vector of data symbols, where $\mathbb{E}\{\mathbf{s}[m]\mathbf{s}^\textsf{H}[m]\} = \frac{\rho}{K}\mathbf{I}_K$, for average power $\rho$ and $m\in \mathcal{M} = \{1,\cdots, M\}$. The data symbols are, then, processed via a $K\times K$ subcarrier-dependent baseband beamformer $\mathbf{F}_\mathrm{BB}[m] = [\mathbf{f}_{\mathrm{BB}_1}[m],\cdots,\mathbf{f}_{\mathrm{BB}_K}[m]]$. In order to steer the generated beams toward users, an  $N\times N_\mathrm{RF}$ subcarrier-independent analog beamformer $\mathbf{F}_\mathrm{RF}$ ($N_\mathrm{RF}=K<{N}$) is employed. Since the analog beamformers are realized with phase-shifters, they have constant-modulus constraint, i.e., $|[\mathbf{F}_\mathrm{RF}]_{i,j}| = {1}/{\sqrt{N}}$ as $i = 1,\cdots, N_\mathrm{RF}$ and $j = 1,\cdots, N$. Then, the transmitted signal, i.e.,  $\mathbf{F}_\mathrm{RF}\mathbf{F}_\mathrm{BB}[m]\mathbf{s}[m]$, is received at the $k$th user  as 
		\begin{align}
		\label{receivedSignal}
	\tilde{	\mathbf{y}}_{k}[m]\hspace{-3pt} =\hspace{-3pt} \mathbf{w}_{\mathrm{RF},k}^\textsf{H}\mathbf{H}_{k}[m]\mathbf{F}_\mathrm{RF}\mathbf{F}_{\mathrm{BB}}[m]\mathbf{s}[m] + \mathbf{w}_{\mathrm{RF},k}^\textsf{H}\mathbf{e}_k[m],
		\end{align}
		where $\mathbf{e}_k\in\mathbb{C}^{\bar{N}}$ is the complex additive white Gaussian noise (AWGN) vector with $\mathbf{e}_k[m] \sim \mathcal{CN}(\mathbf{0},\mathbf{I}_{\bar{N}}\sigma_e^2)$. $\mathbf{w}_{\mathrm{RF},k}\in \mathbb{C}^{\bar{N}}$ denotes the analog beamformer at the receiver with $|[\mathbf{w}_{\mathrm{RF},k}]_n| = 1/\sqrt{\bar{N}}$ for $n \in \{1,\cdots, \bar{N}\}$.}

	Due to limited reflected path components and negligible scattering, the THz channel is usually constructed as the superposition of a single LoS path with a few assisting NLoS paths~\cite{ummimoTareqOverview,elbir2021JointRadarComm,thz_beamSplit}.	In addition, multipath channel models are also widely used, especially for indoor applications~\cite{teraMIMO,ummimoTareqOverview}. Hence, we consider a general scenario, wherein {\color{black}the $ \bar{N}\times N$ channel matrix  is represented by the combination of $L$ paths with physical direction-of-arrival (DOA) and direction-of-departure (DOD) angles $\phi_{k,l} = \sin \tilde{\phi}_{k,l}$ and $\varphi_{k,l} = \sin \tilde{\varphi}_{k,l}$ ($\tilde{\phi}_{k,l}, \tilde{\varphi}_{k,l}\in [-\frac{\pi}{2}, \frac{\pi}{2}]$), respectively~\cite{ummimoTareqOverview}. Due to wideband processing, beam-split occurs and the physical directions ($\phi_{k,l}, \varphi_{k,l}$) are deviated in spatial domain ($\theta_{k,m,l},\vartheta_{k,m,l}$). We define the corresponding subcarrier-dependent spatial directions explicitly as 
		\begin{align}
		\label{physical_spatial_directions}
		\theta_{k,m,l} =  \frac{f_m}{f_c} \phi_{k,l} = \eta_m \phi_{k,l}, 
		\vartheta_{k,m,l} =  \frac{f_m}{f_c} \varphi_{k,l} = \eta_m \varphi_{k,l},
		\end{align}
		where  $f_m = f_c + \frac{B}{M}(m - 1 - \frac{M-1}{2}) $ for  $f_c$ and $B$ being the carrier frequency and bandwidth, respectively. $\eta_m = \frac{f_m}{f_c}$, $c_0$ is speed of light and $d = \frac{c_0}{2f_c}$ is the array element spacing. Now, we can introduce the $\bar{N}\times N$ channel matrix as
		\begin{align} 
		\label{channelModel}
		\mathbf{H}_k[m]  =  
		\zeta  \sum_{l =1}^{L}   \alpha_{k} \bar{\mathbf{a}}(\theta_{k,m,l})  {\mathbf{a}}^\textsf{H}(\vartheta_{k,m,l})  e^{-j2\pi\tau_{k,l} f_m },
		\end{align}
		where $\zeta = \sqrt{{\bar{N}N}/{L}}$,  $\alpha_{k}\in\mathbb{C}$ is the complex path gain and  $\tau_{k,l}$ represents the time delay of the $l$th path corresponding to the array origin. $\bar{\mathbf{a}}(\theta_{k,m,l})\in\mathbb{C}^{\bar{N}}$ and  ${\mathbf{a}}(\vartheta_{k,m,l})\in\mathbb{C}^{{N}}$  are the beam-split-corrupted subcarrier-dependent actual array steering vectors corresponding to spatial DOA/DOD angles. In particular, using \eqref{physical_spatial_directions}, $\bar{\mathbf{a}}(\theta_{k,m,l})$ can be defined in terms of the physical DOA  $\phi_{k,l}$ for uniform linear array (ULA) as
		\begin{align}
		\label{steeringVec1}
		\bar{\mathbf{a}}(\theta_{k,m,l})= [1, e^{-j\pi\eta_m  \phi_{k,l}},\cdots,e^{-j\pi (\bar{N}-1) \eta_m \phi_{k,l}}]^\textsf{T},
		\end{align}
		for which we can define $\mathbf{a}(\vartheta_{k,m,l})$ similarly in terms of $\varphi_{k,l}$.
	}

	\section{Proposed Method}
	
	In mm-Wave systems, the subcarrier-frequencies are close to each other, i.e., $f_{m_1}\approx f_{m_2}, \forall m_{1},m_2 \in \mathcal{M} $, which allows one to  employ a single operating wavelength, i.e., $\lambda_1 = \cdots = \lambda_M = \frac{c_0}{f_c}$, \textcolor{black}{for the design of analog beamformers~\cite{heath2016overview,beamSquintRodriguezFernandez2018Dec,beamSquint_FeiFei_Wang2019Oct}. However, this causes deviations in the direction of the generated beams, especially at the far end subcarriers. Although this deviations might me tolerable in mm-Wave~\cite{heath2016overview,beamSquintRodriguezFernandez2018Dec},	in THz-band, the bandwidth is so wide that a single-wavelength assumption for analog beamforming cannot hold and it leads to the split of physical directions, which are observed as spatial directions in the beamspace.}
	
	\subsection{Beam-Split Model}
	Using (\ref{physical_spatial_directions}) and \eqref{steeringVec1}, we first rewrite the $n$th element of beam-split-corrupted steering vector $\bar{\mathbf{a}}(\theta_{k,m,l})$ as
	\begin{align}
	\label{steeringVec2}
	[\bar{\mathbf{a}}(\theta_{k,m,l})]_n  = e^{-j\pi (n-1)  \theta_{k,m,l}} = e^{-j\pi (n-1)  \eta_m\phi_{k,l}}  ,
	\end{align}
	while the beam-split-free steering vector element is formulated as $[\bar{\mathbf{a}}(\phi_{k,l})]_n = e^{-j\pi (n-1)\phi_{k,l}}$, and we define 
	\begin{align}
	\label{beamSplit2}
	\Delta_{k,l}[m] = \theta_{k,m,l} - \phi_{k,l} = (\eta_m -1)\phi_{k,l},
	\end{align}
	as beam-split to design BSA dictionary in the following.

	\begin{proposition}
		\label{lemma0}
		 Let  $\bar{\mathbf{a}}(\theta_{k,m,l})$ and $\bar{\mathbf{a}}(\phi_{k,l})$ be the  steering vectors corresponding to spatial and physical DOAs, respectively. Then, the linear transformation
		\begin{align}
		\label{beamSplitTransformation}
		\bar{\mathbf{a}}(\theta_{k,m,l}) = \boldsymbol{\Gamma}(\theta_{k,m,l}) \bar{\mathbf{a}}(\phi_{k,l}),
		\end{align}
		holds for  $\boldsymbol{\Gamma}(\theta_{k,m,l})  = \mathrm{diag}\{\boldsymbol{\gamma}(\theta_{k,m,l})\}\in \mathbb{C}^{\bar{N}\times \bar{N}}$, where $\boldsymbol{\gamma}(\theta_{k,m,l}) = [\gamma_{1}(\theta_{k,m,l}),\cdots, \gamma_{\bar{N}}(\theta_{k,m,l})]^\textsf{T}$ and  $ \gamma_{n}(\theta_{k,m,l}) = e^{-j\pi (n-1)\Delta_{k,l}[m]}$.
	\end{proposition}
	\begin{IEEEproof}
		The $n$th element of $\bar{\mathbf{a}}(\theta_{k,m,l})$ and $\bar{\mathbf{a}}(\phi_{k,l})$ can be written as $[\bar{\mathbf{a}}(\theta_{k,m,l})]_n = e^{-j\pi(n-1)\theta_{k,m,l}}$ and $[\bar{\mathbf{a}}(\phi_{k,l})]_n = e^{-j\pi(n-1)\phi_{k,l}}$, respectively.	By using (\ref{beamSplit2}), $\gamma_{n}(\theta_{k,m,l})$ can be obtained from  $\gamma_{n}(\theta_{k,m,l}) = \frac{[\bar{\mathbf{a}}(\theta_{k,m,l})]_n}{[\bar{\mathbf{a}}(\phi_{k,l})]_n}$ as $\gamma_{n}(\theta_{k,m,l}) = e^{-j\pi(n-1)(\theta_{k,m,l}-\phi_{k,l})} = e^{-j\pi(n-1)\Delta_{k,l}[m]}$.
	\end{IEEEproof}
	
	Note that the aforementioned analysis can also be done for $\mathbf{a}(\varphi)$, and the structure of $\boldsymbol{\Gamma}_m(\phi_{k,l})$ allows us to estimated the beam-split at the $m$th subcarrier as discussed below. 
	
	{\color{black}
		\begin{corollary}
			\label{lemma1}
			Given $\boldsymbol{\Gamma}_m(\phi_{k,l})$, the beam-split introduced at the $m$th subcarrier can be uniquely obtained via \cite{elbir_THZ_CE_ArrayPerturbation_Elbir2022Aug}
			\begin{align}
			\label{lemma1BeamSplit}
			\Delta_{k,l}[m] = \frac{1}{\bar{N}-1} \sum_{n = 2}^{\bar{N}} \frac{\angle\{[ \boldsymbol{\gamma}(\theta_{k,m,l})]_n \}}{\pi (n-1)}.
			\end{align}
		\end{corollary}

		\begin{IEEEproof}
			The $n$th entry of $\bar{\mathbf{a}}({\theta}_{k,m,l})$ is given by $	[\bar{\mathbf{a}}({\theta}_{k,m,l})]_n= e^{-j\pi (n-1){\theta}_{k,m,l}  },$
			whose angle is computed as $\Omega_{k,l,n}[m] =\angle [\bar{\mathbf{a}}({\theta}_{k,m,l})]_n=-\pi (n-1){\theta}_{k,m,l}$ for $n = 1,\cdots,\bar{N}$. Next, we compute the unwrapped angles as ${\boldsymbol{\Omega}}_{k,l} [m] = \mathrm{unwrap}\{ [\Omega_{k,l,1}[m],\cdots, \Omega_{k,l,\bar{N}}[m]]^\textsf{T} \}$\footnote{\color{black}The unwrapping operation is necessary to ensure not losing information  due to $[-\pi,\pi]$ periodicity of the exponential (i.e., $|(n-1)\theta_{k,m,l}|> 1$), which is ignored in the previous works~\cite{widebandCE2_GonzalezComa2020Feb,spatialWidebandWang2018May}.}. Using $\boldsymbol{\Omega}_k[m]$, the $n$th element of $\bar{\mathbf{a}}(\phi_{k,l})$  becomes $[\bar{\mathbf{a}}({\theta}_{k,m,l})]_n = e^{-j  \frac{\Omega_{k,l,n}[m]}{\eta_m}  }.$	Substituting $[\bar{\mathbf{a}}({\theta}_{k,m,l})]_n$ into (\ref{beamSplitTransformation}) yields $\gamma_{n}(\theta_{k,m,l})= \frac{[\bar{\mathbf{a}}({\theta}_{k,m,l})]_n}{  	[\bar{\mathbf{a}}({\phi}_{k,l})]_n}$, and  we get
			\begin{align}
			&\angle\{\gamma_{n}(\theta_{k,m,l})\}\hspace{-3pt}=  \hspace{-3pt}\Omega_{k,l,n}[m] - \frac{\Omega_{k,l,n}[m]}{\eta_m}\hspace{-3pt} =\hspace{-3pt} \Omega_{k,l,n}[m] \frac{\eta_m-1}{\eta_m}  \nonumber \\
			&= \pi (n-1) \theta_{k,m,l} (\frac{\theta_{k,m,l} - \phi_{k,l}} {\theta_{k,m,l}} )  = \pi (n-1) \Delta_{k,l}[m].
			\end{align}
			Taking average of $\angle \{\gamma_{n}(\theta_{k,m,l})\}$ for $n = 2,\cdots,\bar{N}$ yields
			\begin{align}
			\label{eqgamma}
			\frac{1}{\bar{N}-1}\sum_{n = 2}^{\bar{N}}\angle\{\gamma_{n}(\theta_{k,m,l})\} = \pi (n-1) \Delta_{k,l}[m].
			\end{align}
			By dividing both sides by \hspace{-2pt}$\pi (n-\hspace{-2pt}1)$ in \eqref{eqgamma}, we get  $\Delta_{k,l}[m]$.
		\end{IEEEproof}
	}

	\subsection{BSA Dictionary Design}
	\label{sec:BSADic}
	{\color{black}
		We design a BSA dictionary whose columns are constructed according to the directions that are affected by beam-split. Hence, for an arbitrary physical directions $\phi_{k,l},\varphi_{k,l}\in [-1,1]$, we can readily compute the spatial directions as $\eta_m\phi_{k,l}$, $\eta_m\varphi_{k,l}$. Using this observation, we design the BSA dictionaries $\bar{\mathcal{C}}_m,\mathcal{C}_m$ composed of steering vectors ${\mathbf{c}}(\theta_m)\in\mathbb{C}^{\bar{N}},\bar{\mathbf{c}}(\vartheta_m)\in\mathbb{C}^{{N}}$ as
		\begin{align}
		\bar{\mathcal{C}}_m &= \{\bar{\mathbf{c}}(\theta_m) |  \bar{\mathbf{c}}(\theta_m)\triangleq \boldsymbol{\Gamma}(\theta_m) \bar{\mathbf{a}}(\phi), \theta_m\in [-\eta_m,\eta_m]   \}, \nonumber\\
		\mathcal{C}_m &= \{\mathbf{c}(\vartheta_m) |  \mathbf{c}(\vartheta_m)\triangleq \boldsymbol{\Gamma}(\vartheta_m) \mathbf{a}(\varphi), \vartheta_m\in [-\eta_m,\eta_m]   \}.\nonumber
		\end{align}
		Using the BSA dictionaries, one can readily obtain the physical directions as $\phi = \theta_m/\eta_m$ and $\varphi = \vartheta_m/\eta_m$, $\forall m\in \mathcal{M}$.
		
	}

	\subsection{Channel Estimation}
	
	In downlink, the channel estimation stage is performed simultaneously by all the users  during channel training. {\color{black}Since the BS employs hybrid beamforming architecture, it activates only a single RF chain in each channel use to transmit the pilot signals during channel acquisition~\cite{heath2016overview,beamSquintRodriguezFernandez2018Dec}. Hence, the BS employs $P$ beamformer vectors as $\tilde{\mathbf{F}}= [\tilde{\mathbf{f}}_1,\cdots, \tilde{\mathbf{f}}_P]\in \mathbb{C}^{N\times P}$ ($|\tilde{\mathbf{f}}_p| = 1/\sqrt{N}$) to send $P$ orthogonal pilot signals, $\tilde{\mathbf{S}} = \mathrm{diag}\{\tilde{s}_1,\cdots, \tilde{s}_P\}\in \mathbb{C}^{P\times P}$. For each pilot, the users employ $\bar{P}$ combining vectors $\tilde{\mathbf{W}} = [\tilde{\mathbf{w}}_1,\cdots, \tilde{\mathbf{w}}_{\bar{P}}]\in \mathbb{C}^{\bar{N}\times \bar{P}}$ ($|\tilde{\mathbf{w}}_p| = 1/\sqrt{\bar{N}}$) in each channel use. Hence, the total channel use for processing all pilots is $P\bar{P}$.} Then, the pilot signals are collected by the $k$th receiver in a $\bar{P}\times P$ matrix as
	\begin{align}
	{\mathbf{Y}}_k[m] = \tilde{\mathbf{W}}^\textsf{H} \mathbf{H}_k[m]\tilde{\mathbf{F}}\tilde{\mathbf{S}} + \tilde{\mathbf{E}}_k[m],
	\end{align}
	where $\tilde{\mathbf{E}}_k[m] = \tilde{\mathbf{W}}^\textsf{H}\mathbf{E}_k[m]$ is the effective noise term. Assuming $\tilde{\mathbf{S}} = \mathbf{I}_P$, we get $	{\mathbf{Y}}_k[m] = \tilde{\mathbf{W}}^\textsf{H}\mathbf{H}_k[m]\tilde{\mathbf{F}}[m] + \tilde{\mathbf{E}}_k[m],$	which can be written in a vector form as
	\begin{align}
	\label{y_vector}
	\mathbf{y}_k[m] = {\mathbf{G}} \mathbf{h}_k[m] + \tilde{\mathbf{e}}_k[m],
	\end{align}
	where $\mathbf{y}_k[m] = \mathrm{vec}\{{\mathbf{Y}}_k[m]\}\in \mathbb{C}^{\bar{P}P}$, ${\mathbf{G}} = \tilde{\mathbf{F}}^\textsf{T}\otimes \tilde{\mathbf{W}}^\textsf{H}\in \mathbb{C}^{\bar{P}P\times \bar{N}N }$, $\mathbf{h}_k[m] = \mathrm{vec}\{\mathbf{H}_k[m]\}$, $\tilde{\mathbf{e}}_k[m] = \mathrm{vec}\{\tilde{\mathbf{E}}_k[m]\}$, where $\otimes$ denotes the Kronecker product. By exploiting the THz channel sparsity, (\ref{y_vector}) is rewritten as
	\begin{align}
	\mathbf{y}_k[m] = \boldsymbol{\Psi}_m  \mathbf{x}_k[m] + \tilde{\mathbf{e}}_k[m],
	\end{align}
	where $\boldsymbol{\Psi}_m = (\tilde{\mathbf{F}}^\textsf{T}\mathbf{C}_m^*) \otimes (\tilde{\mathbf{W}}^\textsf{H} \bar{\mathbf{C}}_m) \in \mathbb{C}^{\bar{P}P\times Q^2}$, for which $\bar{\mathbf{C}}_m = [\bar{\mathbf{c}}(\theta_{m,1}),\cdots,\bar{\mathbf{c}}(\theta_{m,Q}) ]$ and $\mathbf{C}_m = [\mathbf{c}(\vartheta_{m,1}),\cdots,\mathbf{c}(\vartheta_{m,Q}) ]$
	are  the $\bar{N}\times Q$ and ${N\times Q}$ dictionary matrices covering the spatial domain with $\theta_{m,q},\vartheta_{m,q} \in [-\eta_m,\eta_m]$ ($\phi,\varphi \in [-1,1]$) for $q=1,\cdots, Q$ as defined in Sec.~\ref{sec:BSADic}. $\mathbf{x}_k[m]\in\mathbb{C}^{Q^2}$  is an $L$-sparse vector, whose non-zero elements corresponds to the set $\{ z_{k,l}[m]| z_{k,l}[m]\triangleq\zeta \alpha_{k}e^{-j2\pi \tau_{k,l}f_m}, l = 1,\cdots, L \}$.
	
	\textcolor{black}{Using the BSA dictionaries, we can employ OMP algorithm to jointly estimate the THz channel support and the beam-split~\cite{dovelos_THz_CE_channelEstThz2,thz_channelEst_beamsplitPatternDetection_L_Dai}.} The proposed BSA-OMP technique is presented in Algorithm~\ref{alg:BSACE}, wherein the physical angles and beam-split are obtained for $m\in\mathcal{M}$ (Step $3-11$). Then, the estimated THz channel $\hat{\mathbf{H}}_k[m]$ is constructed from the set of steering vectors of physical DOAs (Step $12-17$).

	\begin{algorithm}[t]
		\begin{algorithmic}[1] 
			\caption{ \bf BSA-OMP for channel estimation}
			\color{black}
			\Statex {\textbf{Input:}    $\mathbf{Y}_k[m]$, $\mathbf{G}_m$ and $\eta_m$, $\forall m\in \mathcal{M}$. \label{alg:BSACE}}
			\Statex \textbf{Output:} Channel estimate $\hat{\mathbf{H}}_k[m]$, beam-split $\hat{\Delta}_{k,l}[m]$.
			\State \textbf{for} $k \in\mathcal{K}$
			\State  $l=1$, $\bar{\mathcal{I}}_{l-1} = \mathcal{I}_{l-1} = \emptyset$,
			$\mathbf{r}_{l-1}[m] = \mathbf{y}_k[m], \forall m\in \mathcal{M}$.
			\State \textbf{while} $l\leq L$ \textbf{do}
			
			\State \indent $\{\bar{q}^\star, q^\star \}= \argmax_{\bar{q},q} \sum_{m=1}^{M}|\boldsymbol{\psi}_{\bar{q},q}^\textsf{H}[m]\mathbf{r}_{l-1}[m] |$, \par  \indent where $\hspace{-1pt}\boldsymbol{\psi}_{\bar{q},q}[m] \hspace{-1pt}=\hspace{-1pt} (\tilde{\mathbf{F}}^\textsf{T}\mathbf{c}^*(\theta_{m,q})) \otimes (\tilde{\mathbf{W}}^\textsf{H} \bar{\mathbf{c}}(\vartheta_{m,\bar{q}})) $.
			\State \indent $\bar{\mathcal{I}}_{l} = \bar{\mathcal{I}}_{l-1} \bigcup \{\bar{q}^\star\}$, $\hat{\phi}_{k,l} =\frac{\theta_{m,\bar{q}^\star}}{\eta_m}$.
			\State \indent $\mathcal{I}_{l} = \mathcal{I}_{l-1} \bigcup \{q^\star\}$, $\hat{\varphi}_{k,l} =\frac{\vartheta_{m,q^\star}}{\eta_m}$.
			\State  \indent $\hat{\Delta}_{k,l}[m] = \hat{\theta}_{m,\bar{q}^\star}- \hat{\phi}_{k,l}, \forall m\in \mathcal{M}$.
			\State  \indent $\boldsymbol{\Psi}_m(\mathcal{I}_l) = (\tilde{\mathbf{F}}^\textsf{T}\mathbf{C}_m(\mathcal{I}_l)) \otimes (\tilde{\mathbf{W}}^\textsf{H} \bar{\mathbf{C}}(\bar{\mathcal{I}}_l))$.
			\State  \indent $\mathbf{r}_{l}[m] = \left( \mathbf{I}_{\bar{P}P} -  \boldsymbol{\Psi}_m(\mathcal{I}_l) \boldsymbol{\Psi}_m^\dagger(\mathcal{I}_l) \right) \mathbf{y}_k[m]$.
			\State \indent $l= l + 1$.
			\State \textbf{end while}
			\State   $\bar{\boldsymbol{\Xi}}_k \hspace{-3pt}= \hspace{-3pt} [\bar{\mathbf{a}}(\hat{\phi}_{k,1}), \cdots, \bar{\mathbf{a}}(\hat{\phi}_{k,L})]$. $\boldsymbol{\Xi}_k  \hspace{-3pt}= \hspace{-3pt} [\mathbf{a}(\hat{\varphi}_{k,1}), \cdots, \mathbf{a}(\hat{\varphi}_{k,L})]$.
			\State  \textbf{for} $m\in \mathcal{M}$
			\State \indent$\hat{\mathbf{z}}_k[m] = \boldsymbol{\Psi}_m^\dagger(\mathcal{I}_{l-1}) \mathbf{y}_k[m]$.
			\State \indent$\hat{\mathbf{H}}_k[m] = \bar{\boldsymbol{\Xi}}_k\mathrm{diag}\{\hat{\mathbf{z}}_k[m]\}\boldsymbol{\Xi}_k^\textsf{H}$.
			\State \textbf{end for}
			\State \textbf{end for}
		\end{algorithmic} 
	\end{algorithm}

	{\color{black}
		\subsection{Beamformer Design}
		Once the channel is estimated, it is used to design the hybrid beamformers $\mathbf{F}_\mathrm{RF}$, $\mathbf{F}_\mathrm{BB}[m]$ and $\mathbf{W}_\mathrm{RF} = [\mathbf{w}_{\mathrm{RF},1},\cdots, \mathbf{w}_{\mathrm{RF},K}]$ via maximizing the overall spectral efficiency. The design problem is equivalent to maximizing $ \sum_{m=1}^{M} |\mathbf{u}_{q,\bar{q}}^\textsf{H}[m] \mathbf{v}_k[m]|$, which is the correlation cost between the hybrid  and unconstrained beamformers $\mathbf{F}_\mathrm{opt}[m]\in\mathbb{C}^{N\times K }$ and $\mathbf{W}_\mathrm{opt}[m]\in \mathbb{C}^{\bar{N}\times K}$~\cite{heath2016overview}. Here, $\mathbf{u}_{q,\bar{q}}[m]$ denotes the BSA dictionary columns as 
		$\mathbf{u}_{q,\bar{q}} [m]= [\mathbf{C}_m]_q^* \otimes [\bar{\mathbf{C}}_m]_{\bar{q}} \in \mathbb{C}^{\bar{N}N}$ and $\mathbf{v}_k[m] = \mathbf{f}_{\mathrm{opt},k}^*[m]\otimes \mathbf{w}_{\mathrm{opt},k}[m]\in \mathbb{C}^{\bar{N}N}$ corresponds to the unconstrained beamformers, wherein $\mathbf{f}_{\mathrm{opt},k}[m]$ can be obtained from the singular value decomposition (SVD) of ${\mathbf{H}}_k[m]$~\cite{heath2016overview}, and $\mathbf{w}_{\mathrm{opt},k}[m]=\frac{1}{\rho} \big(\mathbf{f}_{\mathrm{opt},k}^\textsf{H}[m]\mathbf{H}_k^\textsf{H}[m]\mathbf{H}_k[m]\mathbf{f}_{\mathrm{opt},k}[m] + \frac{\sigma_e^2}{\rho}  \big)^{-1}\mathbf{f}_{\mathrm{opt},k}^\textsf{H}[m]\mathbf{H}_k[m]$. Next, we find the $k$th column of the analog beamformers from
		\begin{align}
		\{q^\star,\bar{q}^\star\} = \argmax_{q,\bar{q}} \sum_{m=1}^{M}|\mathbf{u}_{q,\bar{q}}^\textsf{H}[m] \mathbf{v}_k[m] |,
		\end{align}
		where $\{q^\star,\bar{q}^\star\}$ denote the selected beamformer indices. Then, the $k$th column of $\mathbf{F}_\mathrm{RF}$ and $\mathbf{W}_\mathrm{RF}$ are selected from the steering vectors corresponding to the physical directions as $\mathbf{a}(\phi_{q^\star})$ and $ \bar{\mathbf{a}}(\varphi_{\bar{q}^\star})$, respectively. \textcolor{black}{In order to mitigate the inter-user interference, the baseband beamformer is obtained from the $K\times K$ effective channel matrix whose} the $k$th column is  $\left[\mathbf{H}_\mathrm{eff}[m]\right]_k = \mathbf{w}_{\mathrm{RF},k}^\textsf{H}\mathbf{H}_k[m]\mathbf{F}_\mathrm{RF} $, $k\in \mathcal{K}$, and the baseband beamformer is computed as $\mathbf{F}_\mathrm{BB}[m] = \mathbf{H}_\mathrm{eff}[m]^\dagger$. The algorithmic steps of the proposed BSA-OMP approach for beamformer design are presented in Algorithm~\ref{alg:BSAHB}.
	}
	
	{
		\color{black}
		\subsection{Complexity and Training Overhead}
		The  computational complexity of proposed BSA-OMP approaches is the same as traditional OMP techniques~\cite{heath2016overview}. In particular, the complexity of the BSA-OMP approach for channel estimation in Algorithm~\ref{alg:BSACE} is mainly due to the matrix multiplications in step $4$ $(O(M(3P\bar{P} + PN^2 + \bar{P}\bar{N}^2))))$, step $8$ $(O(2P\bar{P}L))$ and step $14$ $(O(P^3\bar{P}^3 + LP^2\bar{P^2} + LP \bar{P}))$. 	The channel training complexity of the proposed BSA-OMP approach requires only $P\bar{P}$ ($16$ times lower, see Sec.~\ref{sec:Sim}) channel use  for pilot signaling while the tradition approaches, e.g., LS, MMSE and ADR~\cite{spatialWidebandWang2018May}, need at least $N\bar{N}$ channel use.
		
		The complexity of the proposed beamforming approach in Algorithm~\ref{alg:BSAHB} is mainly due to operations in steps $3-4$ with total complexity of  $O( K(N+\bar{N})+M N\bar{N}) )$. Furthermore, the proposed approach does not require additional hardware components as opposed to the previous works employing TDNs~\cite{dovelos_THz_CE_channelEstThz2,beamSquint_FeiFei_Wang2019Oct,thz_beamSplit}.

	}
	\begin{algorithm}[t]
		\begin{algorithmic}[1] 
			\caption{ \bf BSA-OMP for hybrid beamforming}
			\color{black}
			\Statex {\textbf{Input:} $\bar{\mathbf{C}}_m$, $\mathbf{C}_m$, $\mathbf{F}_\mathrm{opt}[m]$, $\mathbf{W}_\mathrm{opt}[m]$, $\eta_m$, $m\in \mathcal{M}$. \label{alg:BSAHB}}
			\Statex \textbf{Output:} $\mathbf{F}_\mathrm{RF}$, ${\mathbf{F}}_\mathrm{BB}[m]$, $\mathbf{W}_\mathrm{RF}$.
			\State  $\mathbf{F}_\mathrm{RF} = \emptyset $, $\mathbf{F}_\mathrm{r}[m] =\mathbf{F}_\mathrm{opt}[m]$,\par \hspace{-20pt}$\mathbf{W}_\mathrm{RF} = \emptyset$, $\mathbf{W}_\mathrm{r}[m] =\mathbf{W}_\mathrm{opt}[m]$.
			\State \textbf{for} $k=1,\cdots, N_\mathrm{RF}$ \textbf{do}
			\State \indent  $\mathbf{v}_k[m] =\mathbf{f}_{\mathrm{r},k}^*[m]\otimes \mathbf{w}_{\mathrm{r},k}[m] $, $m \in\mathcal{M}$.
			\State \indent  $\{q^\star,\bar{q}^\star\} = \argmax_{q,\bar{q}} \sum_{m=1}^{M}|\mathbf{u}_{q,\bar{q}}^\textsf{H}[m] \mathbf{v}_k[m] |$.
			\State \indent $\mathbf{F}_\mathrm{RF} =\left[\mathbf{F}_\mathrm{RF}| \mathbf{a}(\phi_{q^\star})  \right]$, $\mathbf{W}_\mathrm{RF} = \left[\mathbf{W}_\mathrm{RF}| \bar{\mathbf{a}}(\varphi_{\bar{q}^\star})  \right]$.
			\State \textbf{end for}
			\State  $\left[\mathbf{H}_\mathrm{eff}[m]\right]_k = \mathbf{w}_{\mathrm{RF},k}^\textsf{H}\mathbf{H}_k[m]\mathbf{F}_\mathrm{RF} $, $k\in \mathcal{K}$.
			\State  $\mathbf{F}_\mathrm{BB}[m] = \mathbf{H}_\mathrm{eff}[m]^\dagger$, $m\in \mathcal{M}$.
			
			\State  $\left[\mathbf{F}_\mathrm{BB}[m]\right]_k = \left[\mathbf{F}_\mathrm{BB}[m]\right]_k / \| \mathbf{F}_\mathrm{RF}\mathbf{F}_\mathrm{BB}[m] \|_\mathcal{F}$, $k \in \mathcal{K}$.
		\end{algorithmic} 
	\end{algorithm}
	
	\begin{figure}[t]
		\centering
		{\includegraphics[draft=false,width=\columnwidth]{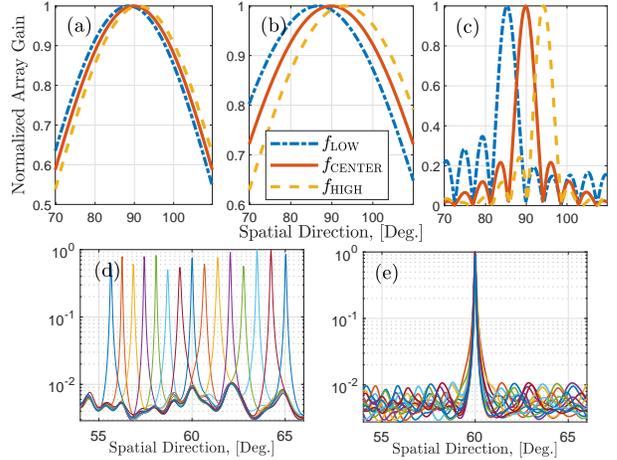} } 
		\caption{Array gain with respect to spatial direction at low, center and high end subcarriers for (a) $f_c=3.5$ GHz, $B=0.1$ GHz; (b) $f_c=28$ GHz, $B=2$ GHz; and (c) $f_c=300$ GHz, $B=30$ GHz, respectively. Array gain corresponding to the all subcarriers (d) under the effect of beam-split and (e) the beam-split-corrected array gain by using BSA-OMP.
		}
		\label{fig_ArrayGain}
	\end{figure}

	\section{Numerical Experiments}
	\label{sec:Sim}
	Throughout the simulations (unless stated otherwise), the signal model in (\ref{channelModel}) is generated with $f_c=300$ GHz, $B=30$ GHz, $M=128$, $N= 256$, $\bar{N}=16$, $P=\bar{P}=16$,  $N_\mathrm{RF}=8$, $L=3$, $K=8$, $Q=8N$~\cite{ummimoTareqOverview,thz_beamSplit}. 	The beamformer matrices $\tilde{\mathbf{F}}$ and $\tilde{\mathbf{W}}$ are modeled with random phases as  $[\tilde{\mathbf{F}}]_{i,j} \frac{1}{\sqrt{N}}e^{j{\psi_F}}$, $[\tilde{\mathbf{W}}]_{i,j} \frac{1}{\sqrt{\bar{N}}}e^{j{\psi_W}}$, where $\psi_F, \psi_W \sim \mathrm{uniform}[-\frac{\pi}{2},\frac{\pi}{2}]$.
	
	Fig.~\ref{fig_ArrayGain} shows the array gain with respect to physical DOA of a single user scenario. In Fig.~\ref{fig_ArrayGain}(a)-(b), array gain is computed for the user physical DOA at $30^\circ$ and $60^\circ$ for all subcarriers when $M=15$. We can see significant deviation in the beamspace due to  beam-split, which is about $3^\circ$ ($6^\circ$) for the physical DOA $30^\circ$ ($60^\circ$), respectively. The array gain computed by using the proposed BSA approach is also given in Fig.~\ref{fig_ArrayGain}(c)-(d), respectively. As it is seen, the proposed BSA approach effectively mitigates the impact of beam-split and the spectra of all subcarriers align toward the same (physical channel) direction. 
	
\begin{figure}[t]
		\centering
		{\includegraphics[draft=false,width=.9\columnwidth]{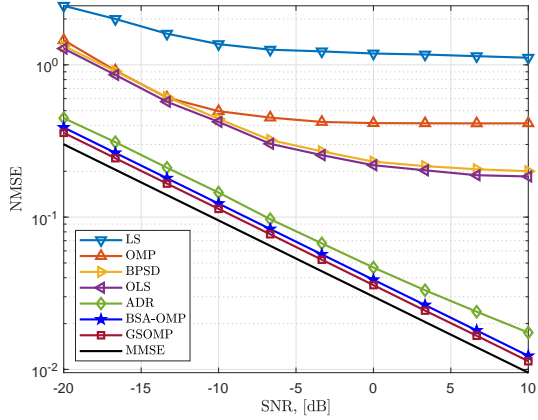} } 
		\caption{THz wideband channel estimation, NMSE versus SNR.
		}
		\label{fig_ChannelEst}
	\end{figure}

	In Fig.~\ref{fig_ChannelEst}, THz channel estimation performance of the proposed BSA-OMP approach is evaluated, in comparison with the state-of-the-art techniques, OMP~\cite{beamSquintRodriguezFernandez2018Dec}, least-squares (LS), oracle LS (OLS), minimum mean-squared-error (MMSE), generalized simultaneous OMP (GSOMP)~\cite{dovelos_THz_CE_channelEstThz2}, \textcolor{black}{ADR~\cite{spatialWidebandWang2018May}}, BSPD~\cite{thz_channelEst_beamsplitPatternDetection_L_Dai}, in terms of normalized MSE of the channel: $\mathrm{NMSE} = \frac{||\mathbf{h}_k[m] - \hat{\mathbf{h}}_k[m] ||_2^2}{||\mathbf{h}_k[m] ||_2^2}$ against signal-to-noise ratio (SNR). \textcolor{black}{Note that the MMSE estimator relies on the true covariance matrix and it is computed for each subcarrier exclusively, hence it is regarded as the benchmark method in this work.} We can see from Fig.~\ref{fig_ChannelEst} that poor performance is achieved by the conventional techniques, e.g., OMP, LS and OLS, that do not take into account the effect of beam-split. \textcolor{black}{Although ADR employs an angular rotation to mitigate the beam-split, it encounters error floors at high SNR regions due the use of coarse beam-split estimation and low resolution Fourier dictionary.} Similarly, BSPD uses a beam-split detection mechanism, it fails to recover the channel supports accurately. On the other hand, the proposed BSA-OMP approach achieves much better performance compared to aforementioned techniques and attain a close performance to MMSE and GSOMP, which employs additional hardware components, i.e., TDNs, to mitigate beam-split. In contrast, our BSA-OMP approach does not require TDN, hence, it is more hardware-efficient while maintaining sufficient NMSE performance.

\begin{figure}[t]
		\centering
		{\includegraphics[draft=false,width=.9\columnwidth]{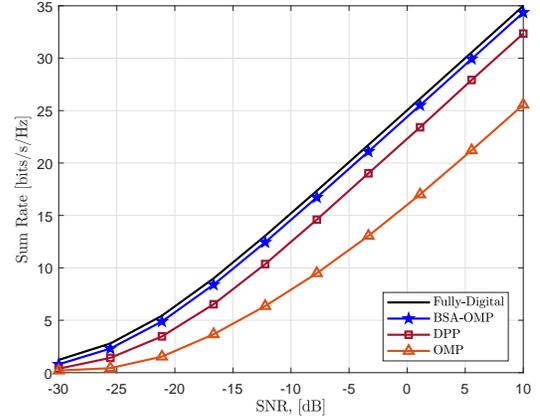} } 
		\caption{THz wideband hybrid beamforming, sum-rate versus SNR.
		}
		\label{fig_Beamforming}
	\end{figure}

	{\color{black}
		Fig.\ref{fig_Beamforming} shows the hybrid beamforming performance in terms of sum-rate, wherein the unconstrained beamformers, i.e., $\mathbf{F}_\mathrm{opt}[m]$, $\mathbf{W}_\mathrm{opt}[m]$, are employed for fully-digital beamformer as a benchmark. A significant performance improvement is observed when the proposed BSA approach is applied to the traditional OMP technique~\cite{heath2016overview}, which does not take into account the effect of beam-split. The proposed method is also superior than the delay-phase precoding (DPP) method~\cite{delayPhasePrecoding_THz_Dai2022Mar}, which employs a TDN to realize frequency-dependent analog beamformer while no such requirement is needed for BSA-OMP.
	}

	\section{Summary}
	{\color{black}In this work, we introduced a BSA-OMP approach to effectively mitigate the effect of beam-split for THz channel estimation and beamforming. Compared to prior works, the  proposed approach is advantageous since it does not require additional hardware components, e.g., TDNs, while maintaining close-to-MMSE (fully-digital) channel estimation (beamforming) performance. Furthermore, the complexity of BSA-OMP is low and the same as for the traditional OMP technique.}

	%

	\bibliographystyle{IEEEtran}
	\bibliography{IEEEabrv,references_107}

	%
	%
	%
	%
	%
	%
	%
	%
	

	%

\end{document}